\documentclass[a4paper]{jpconf}
\usepackage{graphicx,amsmath,bm}
\usepackage{citesort}

\newcommand{\Ft}{\,^*\! F}
\newcommand{\ft}{\,^*\! f}

\newcommand{\vv}[1]{\boldsymbol{#1}}

\newcommand{\beq}{\begin{equation}}
\newcommand{\eeq}{\end{equation}}

\begin{document}
\title{An operator-splitting numerical scheme for relativistic magnetohydrodynamics}

\author{David Phillips and Serguei Komissarov}

\address{School of Mathematics, University of Leeds, UK}

\ead{mmdnhp@leeds.ac.uk, s.s.komissarov@leeds.ac.uk}

\begin{abstract} 
We describe a novel operator-splitting approach to numerical relativistic magnetohydrodynamics designed to expand its applicability to the domain of  ultra-high magnetisation. In this approach, the electromagnetic field is split into the force-free component, governed by the equations of force-free degenerate electrodynamics (FFDE), and the perturbation component, governed by the perturbation equations derived from the full system of relativistic magnetohydrodynamics (RMHD). The combined system of the FFDE and perturbation equations is integrated simultaneously, for which various numerical techniques developed for hyperbolic conservation laws can be used. At the end of every time-step of numerical integration, the force-free and the perturbation components of the electromagnetic field are recombined and the result is regarded as the initial value of the force-free component for the next time-step, whereas the initial value of the perturbation component is set to zero.   To explore the potential of this approach, we build a 3rd-order WENO code, which was used to carry out 1D and 2D test simulations.  Their results show that this operator-splitting  approach allows us to bypass the stiffness of RMHD in the ultra-high-magnetisation regime where the perturbation component becomes very small.  At the same time, the code performs very well in problems with moderate and low magnetisation too.    
\end{abstract}

\section{Introduction}
The strong gravity of astrophysical black holes and neutron stars creates some of the most extreme physical conditions in the Universe, which cannot be achieved in research laboratories. In particular, they naturally develop magnetospheres with extremely high plasma magnetisation. In contrast to the non-relativistic  problems, where the magnetisation is well described by the ratio of thermodynamic and magnetic  pressures ($\beta=p/p_m$), the magnetisation of relativistic plasma is best described by the parameter $\sigma=\hat{B}^2/4\pi (e+p)$, where $\hat{B}$ is the magnetic field strength as measured in the rest frame of plasma, $p$ is the thermal pressure and $e=\rho c^2 + e_t$, where $\rho$ and $e_t$ are the proper rest-mass density and thermal energy of plasma, respectively.  In the magnetospheres of neutron start $\sigma$ can reach the values of the order of $10^3-10^6$.  Unfortunately, modern conservative schemes for relativistic magnetohydrodynamics (RMHD) begin to fail in multi-dimensional problems already when $\sigma$ is of the order of few. Namely, the conservative variables cannot be converted to physically meaningful primitive variables.  Higher values of $\sigma$ can be handled for isentropic flows, where the energy equation can be eliminated \cite{SSK07}. A fix based on the entropy obtained via parallel integration of the adiabatic entropy transport equation can help to bypass crashes for adiabatic flows as well \cite{Noble09}.  Obviously, this approach would not work for problems involving shocks, like the termination shock of pulsar winds \cite{Porth13}, and, more importantly, magnetic reconnection, which involves kinetic and magnetic energy dissipation. Excessive artificial plasma heating due to numerical resistivity is another issue in simulations of highly magnetised flows. 

In the limit of infinite magnetisation, $\sigma\to\infty$, the system of RMHD becomes degenerate. It reduces to the Force-Free Degenerate Electrodynamics, where only two components of the energy-momentum equation are independent \cite{SSK02}.  It has been suggested, that this is the main reason for the shortcomings of conservative schemes in high-magnetisation problems \cite{SSK06}.   If so, it would make sense to apply a perturbation approach, where the FFDE solution is a leading order approximation and the plasma inertia enters only equations governing small perturbations to this solution.        

A similar issue have been identified some time ago in Newtonian MHD simulations of the collision between highly magnetised ($\beta\ll1$) Earth's magnetosphere and the solar wind. In that problem, the Earth's largely dipolar magnetic field increases by many orders of magnitude from the collision site to the troposphere, whereas the perturbation of this field remains of about the same magnitude and hence increasingly small relative to the dipolar field on approach to the troposphere. As a result, the perturbation suffers large computational errors.   To overcome this problem, Tanaka \cite{Tanaka94} proposed to separate the strong stationary dipolar field from its perturbation and numerically integrate only the nonlinear equations governing the perturbation. This approach has proven very productive \cite{Eggington20}.  Here we describe a generalisation of this approach, where the zero-order solution is time-dependent.

\section{Ideal relativistic magnetohydrodynamics}

The covariant system of ideal RMHD consists of the Faraday equation

\begin{equation}
  \nabla_\beta  \Ft^{\alpha \beta} = 0 \, ,
\label{Maxw1}
\end{equation}
the energy-momentum equation 

\begin{equation}
     \nabla_\nu T^{\mu\nu} =0 \, ,
\label{EMc}
\end{equation}
the continuity equation 

\begin{equation}
   \nabla_\nu {\rho u^\nu}=0 \, ,
\end{equation}
and the perfect conductivity condition

\begin{equation}
    F_{\mu\nu} u^\nu=0 \, .
\end{equation}
In these equations, $F_{\mu\nu}$ is the Maxwell tensor and $\Ft_{\mu\nu}$ is
the Faraday tensor, $T^{\nu\mu}$ is the total stress-energy-momentum tensor, 
$\rho$ is the rest mass energy of plasma particles, and $u^\mu$ is the 
four-velocity vector of macroscopic motion. In highly ionised plasma the 
Faraday tensor is Hodge dual to the Maxwell tensor, and hence  
\begin{equation}
\Ft^{\alpha  \beta} = \frac{1}{2} e^{\alpha \beta \mu \nu} F_{\mu \nu}\,,
\quad 
F^{\alpha  \beta} = -\frac{1}{2} e^{\alpha \beta \mu \nu} \Ft_{\mu \nu},
\label{F}
\end{equation}
where $e_{\alpha \beta \mu \nu} = \sqrt{-g}\,\epsilon_{\alpha \beta \mu \nu}$
is the Levi-Civita alternating tensor of space-time and  $\epsilon_{\alpha \beta \mu \nu}$ is the four-dimensional Levi-Civita symbol. The total stress-energy momentum tensor,

\begin{equation}
     T^{\mu\nu} = T_{(m)}^{\mu\nu} + T_{(e)}^{\mu\nu}\, ,
\label{tsem-1}
\end{equation}
is the sum of the stress-energy momentum tensor of electromagnetic field

\begin{equation}
   T_{(e)}^{\mu\nu} = F^{\mu\gamma} F^\nu_{\ \gamma} -
   \frac{1}{4}(F^{\alpha\beta}F_{\alpha\beta})g^{\mu\nu}\, ,
\label{semt-e}
\end{equation}
and the stress-energy momentum tensor of matter

\begin{equation}
   T_{(m)}^{\mu\nu} = wu^\mu u^\nu + p g^{\mu\nu}\, .
\label{semt-m}
\end{equation}
Here $p$ is the thermodynamic pressure, $w(p,\rho)$ is the relativistic enthalpy per
unit volume as measured in the rest frame of fluid, and $g^{\mu\nu}$ is the metric tensor 
of space-time.

\section{Force-free and perturbation equations}

Let us split the electromagnetic field into the force-free part  $f^{\mu\nu}_{(0)}$ and the correction part $f^{\mu\nu}_{(1)}$
\beq
  F^{\mu\nu} = f^{\mu\nu}_{(0)}+f^{\mu\nu}_{(1)} \,.
\label{exp-F}
\eeq    
The corresponding expansion for the stress-energy-momentum tensor of the 
electromagnetic field is 
\beq
   T^{\mu\nu}_{(e)} = t^{\mu\nu}_{(0)}+ t^{\mu\nu}_{(1)} \,,
\eeq
where 
\beq
   t^{\mu\nu}_{(0)}= f^{\mu\gamma}_{(0)} f^\nu_{(0) \gamma} -
   \frac{1}{4}(f^{\alpha\beta}_{(0)}f_{\alpha\beta}^{(0)})g^{\mu\nu} 
\label{FF-EMT}
\eeq
and 
\beq
   t^{\mu\nu}_{(1)}= 
   f^{\mu\gamma}_{(0)} f^\nu_{(1) \gamma} + f^{\mu\gamma}_{(1)} f^\nu_{(0) \gamma} + 
   f^{\mu\gamma}_{(1)} f^\nu_{(1) \gamma}  -
   \frac{1}{4} \left( 2f^{\alpha\beta}_{(0)}f_{\alpha\beta}^{(1)}  + 
              f^{\alpha\beta}_{(1)}f_{\alpha\beta}^{(1)} \right)g^{\mu\nu} \, .  
\label{PT-EMT}
\eeq
The differential equations of the leading order system are the FFDE equations
\begin{equation}
  \nabla_\beta  \ft^{\alpha \beta}_{(0)} = 0 \, ,
\label{FF-FE}
\end{equation}
\begin{equation}
     \nabla_\nu t^{\mu\nu}_{(0)} =0 \, .
\label{FF-EME}
\end{equation}
Since the four velocity of plasma does not enter these equations, the perfect conductivity condition is written in the form 
\begin{equation}
      \ft^{\mu\nu}_{(0)} f_{\mu\nu}^{(0)} =0 \,,\quad 
      f^{\mu\nu}_{(0)} f_{\mu\nu}^{(0)} >0\,, 
\end{equation}
which ensures the existence of inertial frames where the electric field vanishes. 

The perturbation equations equations are obtained from the the full system of RMHD by removing the terms vanishing due to Eqs.\ref{FF-FE},\ref{FF-EME}   

\begin{equation}
  \nabla_\beta  \ft^{\alpha \beta}_{(1)} = 0 \, ,
  \label{PE-FE}
\end{equation}

\begin{equation}
     \nabla_\nu (t^{\mu\nu}_{(1)}+ T_{(m)}^{\mu\nu} ) =0 \, ,
  \label{PE-EME}
\end{equation}

\begin{equation}
   \nabla_\nu \rho u^\nu=0 \, . 
    \label{PE-CE}
\end{equation}
It is important that this system is fully nonlinear with respect to its hydrodynamic  
component and hence allows the non-linear steepening of magnetosonic waves. 
In order to close this system we use the perfect conductivity condition

\beq
     (f^{\mu\nu}_{(0)}+f^{\mu\nu}_{(1)})u_\mu = 0 \,. 
\eeq
By construction, the perturbation equations do not involve terms quadratic in the leading order electromagnetic field, like $f^{\mu\gamma}_{(0)} f^\nu_{(0) \gamma} $, which would be dominant in problems with high $\sigma$. This is the main goal behind our approach. The terms linear in $f^{\mu\gamma}_{(0)}$ in the perturbation stress-energy-momentum  tensor (\ref{PT-EMT}) describe the interaction with the FFDE field. 

\section{Numerical scheme}

Since the FFDE equations are independent from the perturbation equations, even if initially the electromagnetic component of RMHD solution is close to the FFDE solution, they will eventually diverge.  Once the solutions have diverged, the perturbation of the electromagnetic field is no longer small and the splitting becomes useless. To avoid this issue, the leading order and perturbation components of the electromagnetic field are recombined at the end of every full integration time-step, $F^{\mu\nu} = f^{\mu\nu}_{(0)}+f^{\mu\nu}_{(1)}$ and split at the beginning of the next time-step as  $f^{\mu\nu}_{(0)}=F^{\mu\nu}$,  $f^{\mu\nu}_{(1)}=0$. In other words,  the electromagnetic field is first advanced as force-free and then corrected for the interaction with plasma, making our approach a variant of the operator-splitting method.  

We have explored the potential of this approach using a 3rd order WENO scheme for Special Relativistic MHD,  similar to the one described in \cite{DelZanna07}. 
In this explicit scheme, the solution is advanced forward in time using a 3rd order Runge-Kutta method.  The FFDE equations \ref{FF-FE}-\ref{FF-EME} and the perturbation equations \ref{PE-FE}-\ref{PE-CE}  are integrated simultaneously, without recombination of the electromagnetic field at the Runge-Kutta sub-steps. 
To set up the Riemann problems at the cell interfaces, we developed a novel WENO-interpolation which ensures rapid onset of 3rd convergence even for smooth solutions with local extrema. To compute the interface fluxes, we used an HLL Riemann solver \cite{HLL}, with the speed of light as the fastest wave speeds for the FFDE sub-system, and the fast-magnetosonic speeds for the perturbation sub-system.   The conserved variables of the FFDE are converted into its primitive variables as in \cite{SSK02}, where the energy equation is not utilised at all. The conserved variables of the perturbation sub-system are converted into its primitive variables via the approach similar to the one described in  \cite{DelZanna07}.   Full details of the scheme will be presented elsewhere.    

\section{Test simulations} 
Here, we present the results of some test simulations. We use relativistic units where $c=1$ and $4\pi$ does not appear in Maxwell equations, and the equation of state perfect gas $w=\rho+\Gamma p/(\Gamma-1)$ with the ratio of specific heats $\Gamma=4/3$.  

\begin{figure}[h]
\includegraphics[width=0.5\textwidth]{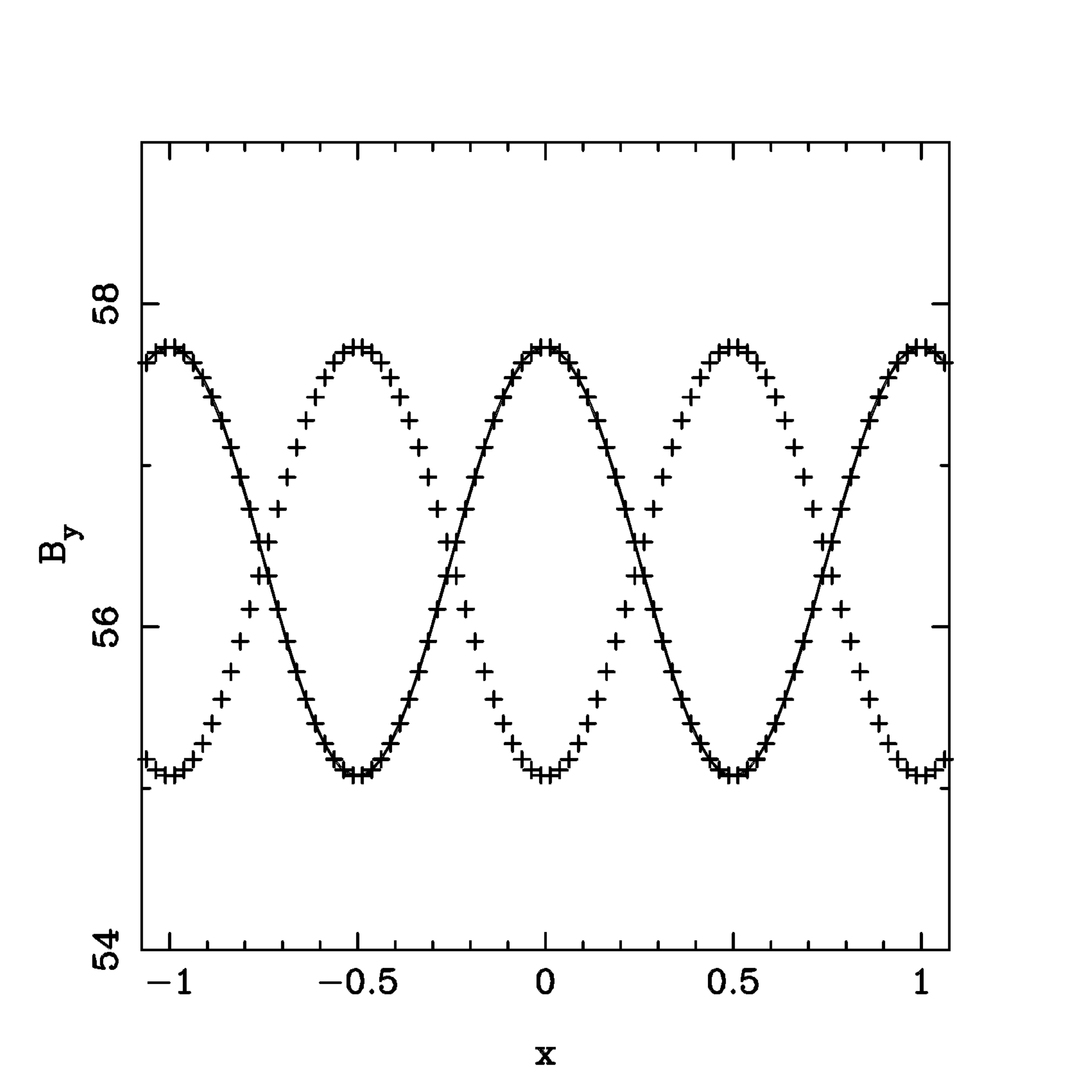}
\includegraphics[width=0.5\textwidth]{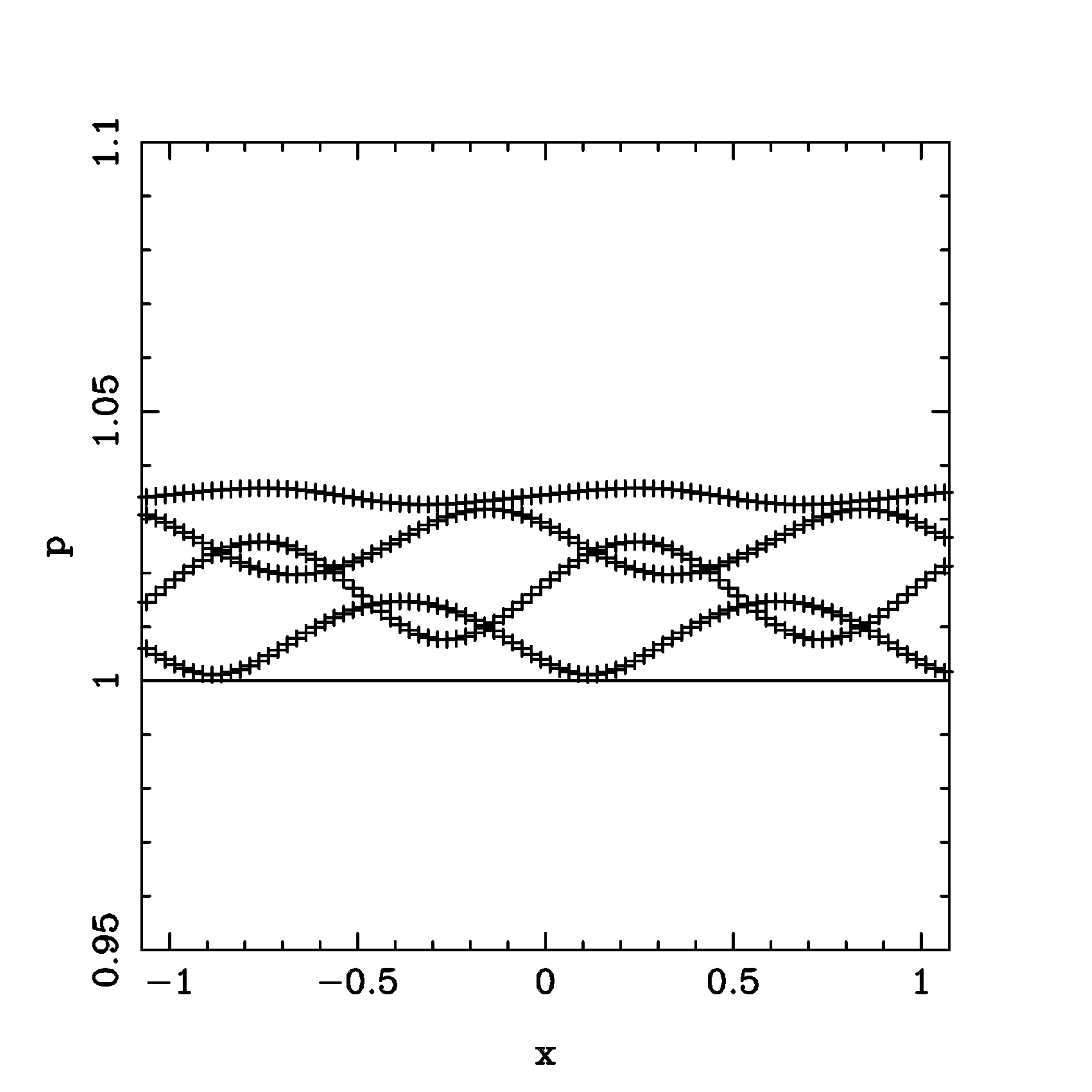}
\caption{Alfv\'en wave test. In the exact solution, the wave moves to the left with the speed $v_a=0.5$.  On both panels, the solid line shows the initial solution ($t=0$) and markers show the solutions for $t=1,2,3,4$ (In the $B_y$-plots, the solutions for $t=1,3$ and $t=0,2,4$ are indistinguishable.).  The resolution is n=80 and the Courant number $Cu=0.4$.}
\label{fig:AW}
\end{figure}

\subsection{1D periodic Alfv\'en wave}
The analytic solution to Alfv\'en waves of finite amplitude is given in \cite{SSK97}. In the exact solution used for the test problem, $\rho(x,t)=p(x,t)=1$, $\sigma(x,t)=\sigma_0$, $B_x=0.3B_0$ and the tangential magnetic field $B_t =\gamma_a B_0 e^{i\psi(x,t)}$, with $\psi(x,t)=\arcsin(0.3\sin(\pi(x+v_a t)))$, where $\gamma_a=(1-v_a^2)^{-1/2}$ and the wavespeed $v_a=0.5$.  $B_0$ is the parameter determining the magnetic field strength in the Hoffmann-Teller frame, it is equivalent to the magnetisation parameter $\sigma_0$.  Figure 1 illustrates the results for $\sigma_0=545$ and table \ref{tab:convergence} shows the results of the convergence study based on this problem, confirming the 3rd order accuracy of the scheme.   For lower $\sigma_0$, the results are almost as good. The relative $L_1$ error for $B_y$ increases only by the factor of about 4.5 for $\sigma_0=5.45\times 10^{-4}$,  suggesting that the approach is suitable for both the high- and low-$\sigma$ regimes.

\begin{table}[h]
\label{tab:convergence}
\caption{Convergence study based on the 1D Alfv\'en wave test. The numerical and analytic solutions  are compared at $t=4$.} 

\begin{center}
\lineup
\begin{tabular}{*{6}{c}}
\br                              
$\0\0n$& 20 & 40 & 80 & 160 & 320 \cr 
\mr
$L_1(B_y)$ &7.48E-2  & 8.89E-3 & 1.03E-3 & 1.31E-4  & 1.64E-5 \cr
$L_1(\rho)$ & 5.83E-1 & 1.11E-1 & 1.51E-2 & 1.90E-3 & 2.38E-4 \cr 

\br
\end{tabular}
\end{center}
\end{table}

\begin{figure}[h]
\includegraphics[width=0.5\textwidth]{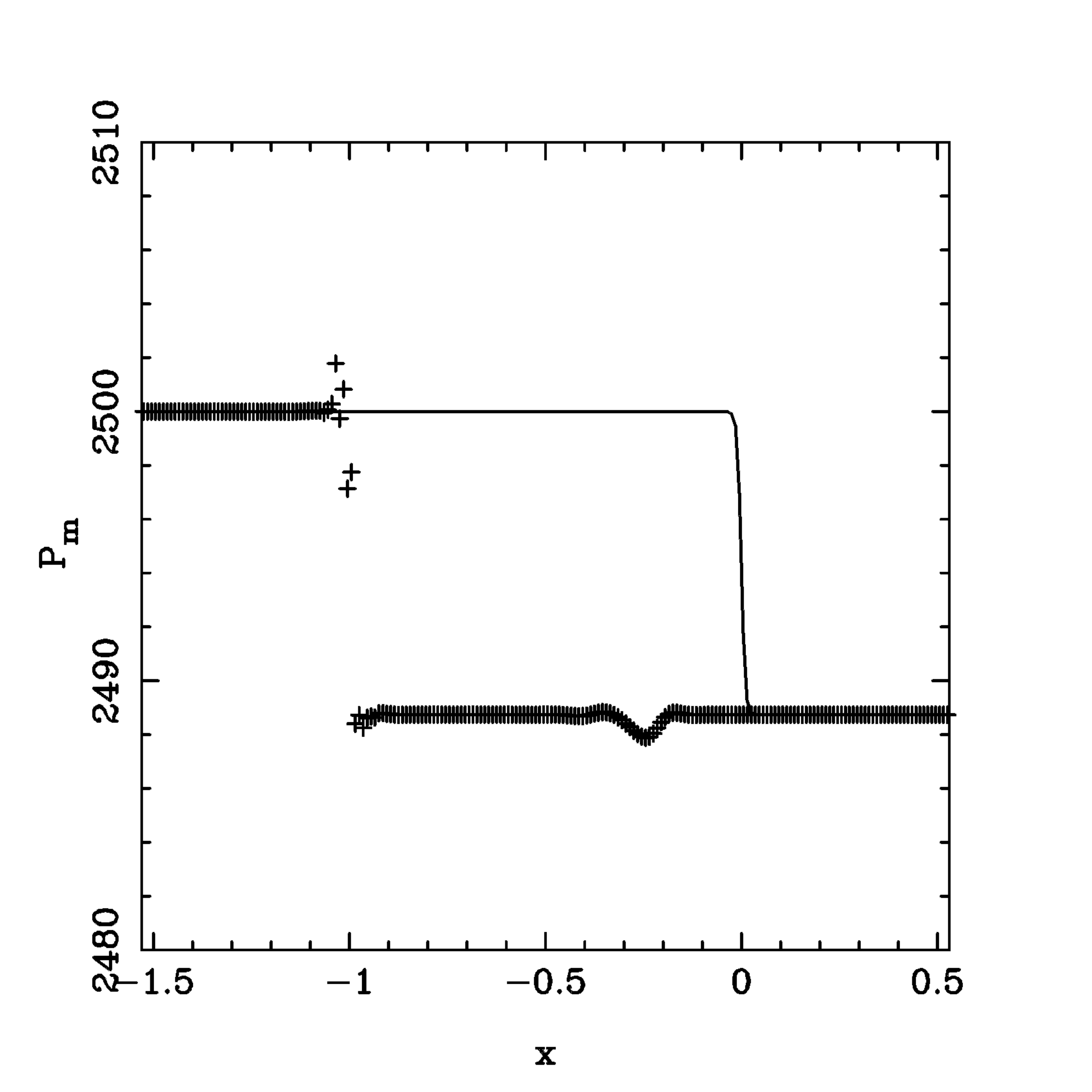}
\includegraphics[width=0.5\textwidth]{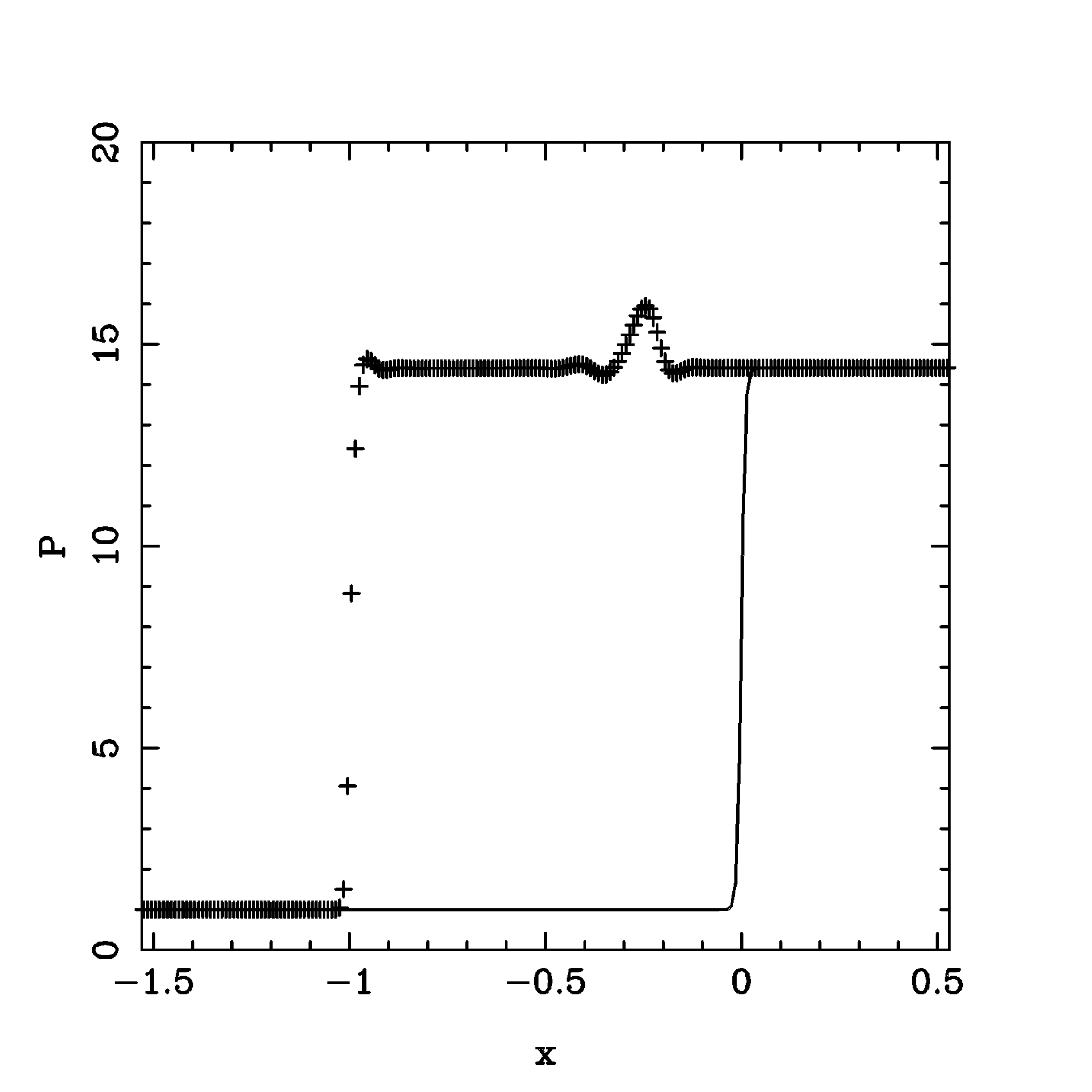}
\caption{\label{fig:SS}1D slow shock test. The magnetic pressure (left panel) and 
the gas pressure (right panel) are shown for the initial solution ($t=0$, solid lines) and the numerical solution at $t=2$ (markers). The exact shock speed is $v_s=0.5$. The disturbance located near $x=-0.3$ at $t=2$ is caused by the evolution of the numerical shock structure at the beginning of the simulations.}
\end{figure}

\subsection{1D slow shock} 
The exact solutions for this and other shock tests were obtained using the method described in \cite{MA87}. Left state: $p=\rho=1$, $\vv{B}=(50,51.03,0)$, $\vv{v}=(0.1995,0,0)$, $\sigma=1000$; Right state: $p=14.41$, $\rho=5.879$, $\vv{B}=(50,50.82,0)$, $\vv{v}=(-0.42.12,-0.6338,0)$, $\sigma=461$. The shock moves to the left with the speed $v_s=0.5$. The angle between the shock front and magnetic field in the rest frame of the upstream state is $45^o$. The results at $t=2$ are illustrated in figure \ref{fig:SS}.

\subsection{1D fast shocks} 

Here we present two tests, one for the high-$\sigma$ regime and one for the low-$\sigma$ regime. In both these cases, the angle between the shock front and the magnetic field in the rest frame of the upstream state is $45^o$. The fast magnetosonic Mach number of the shock in the same frame is $M_f=2$, and relative to the grid the shock moves to the left with the speed $v_s=0.5$. In the initial solution ($t=0$) the shock is located at $x=0$.

\begin{figure}[h]
\includegraphics[width=0.5\textwidth]{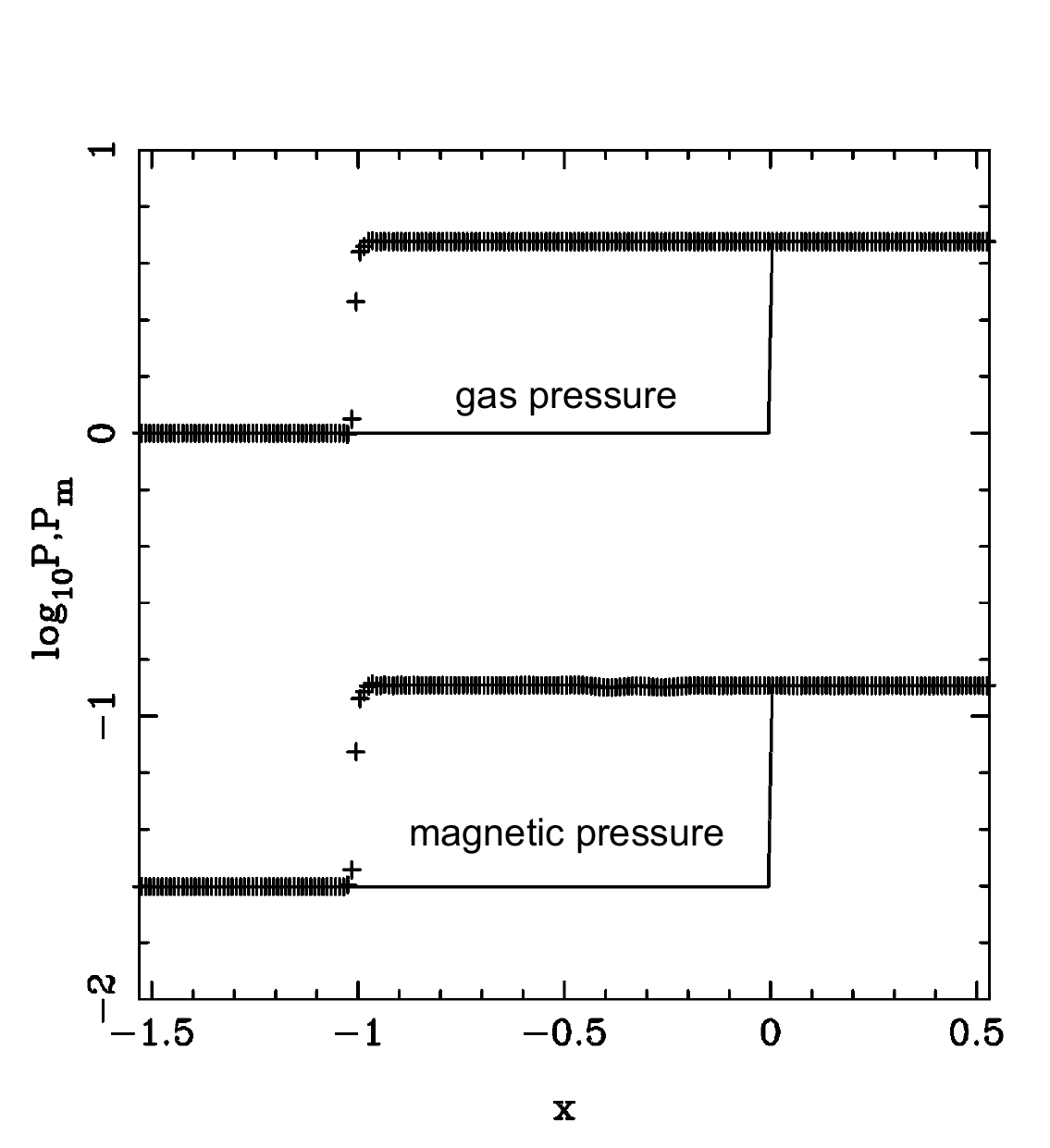}
\includegraphics[width=0.5\textwidth]{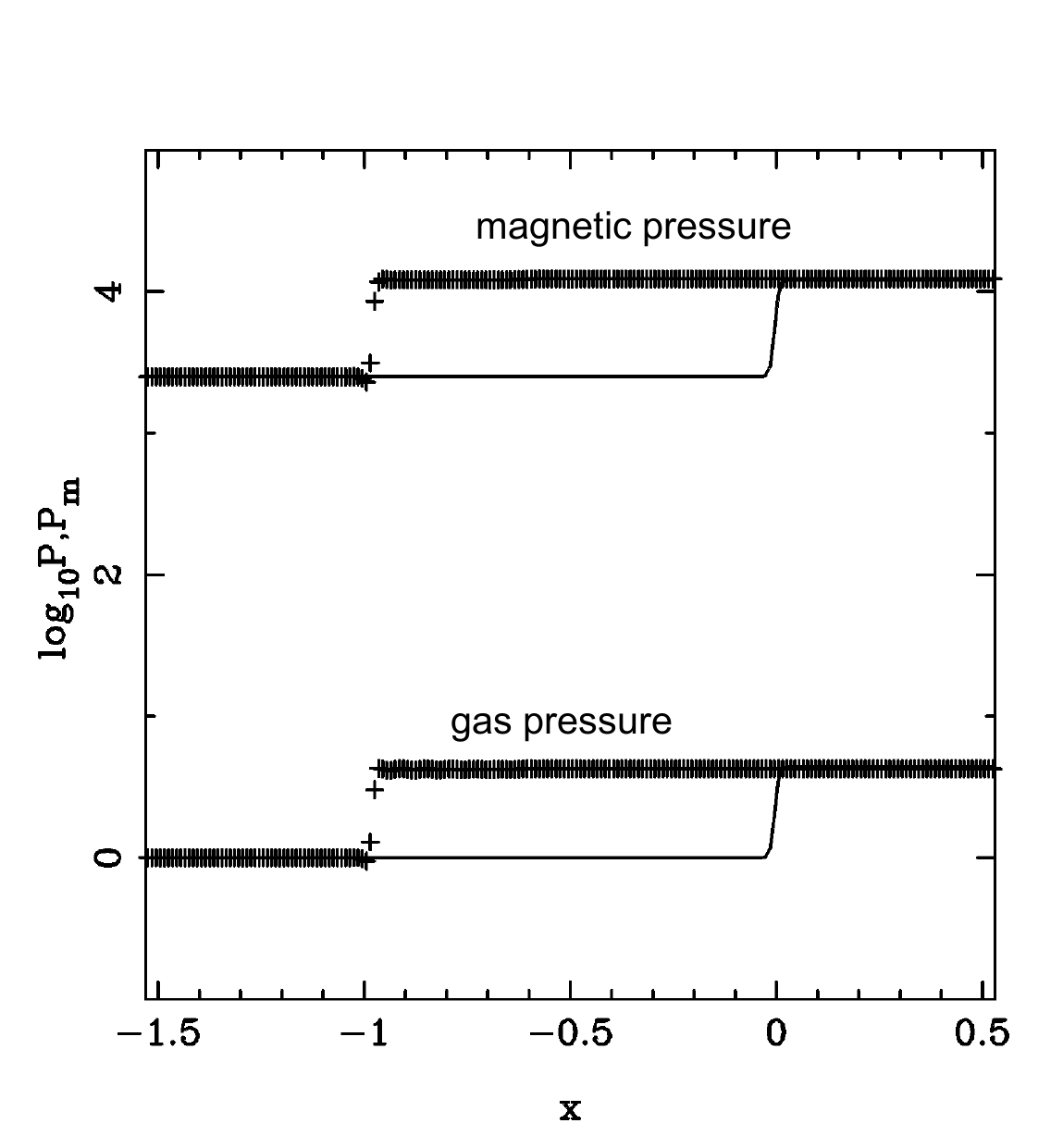}
\caption{\label{fig:FS} Low-$\sigma$ (left panel) and high-$\sigma$ (right panel) fast shock tests. The magnetic pressure (left panel) and the gas pressure (right panel) are shown for the initial solution ($t=0$, solid lines) and for the numerical solution at $t=2$ (markers). The exact shock speed is $v_s=0.5$ in both the cases. }
\end{figure}

{\it High-$\sigma$ case.} Left state:  $p=\rho=1$, $\vv{B}=(50,1985.,0)$, $\vv{v}=(0.9997,0,0)$, $\sigma=1000$. Right state: $p=4.424$, $\rho=2.618$, $\vv{B}=(50,1989.,0)$, $\vv{v}=(0.9977,0.0172,0)$, $\sigma=1119$. The numerical solution at $t=2$ is illustrated in the right panel of figure \ref{fig:FS}.

{\it Low-$\sigma$ case.} Left state:  $p=\rho=1$, $\vv{B}=(0.1581,0.1765,0)$, $\vv{v}=(0.4445,0,0)$, $\sigma=0.01$. Right state: $p=4.754$, $\rho=3.032$, $\vv{B}=(0.1581,0.4876,0)$, $\vv{v}=(-0.1566,0.0048,0)$, $\sigma=0.0116$. The numerical solution at $t=2$ is illustrated in the left panel of figure \ref{fig:FS}.

Shock problems are the most difficult for any numerical scheme.  The lack of energy conservation in the FFDE sub-system is a particular source of concern specific to our scheme.  The presented tests, however, demonstrate that the code can handle at least some shocks quite well. These and other tests, show good performance for shocks characterised by small variation of the tangential magnetic field. However, when this variation becomes strong the computational errors grow and the code may even crash. One example of these problematic shocks is the above high-$\sigma$ fast shock when set up in the rest frame of the upstream state. However, the issue may not be as severe as it seems, as illustrated by the strong explosion test considered next.

\begin{figure}[h]
\includegraphics[width=0.5\textwidth]{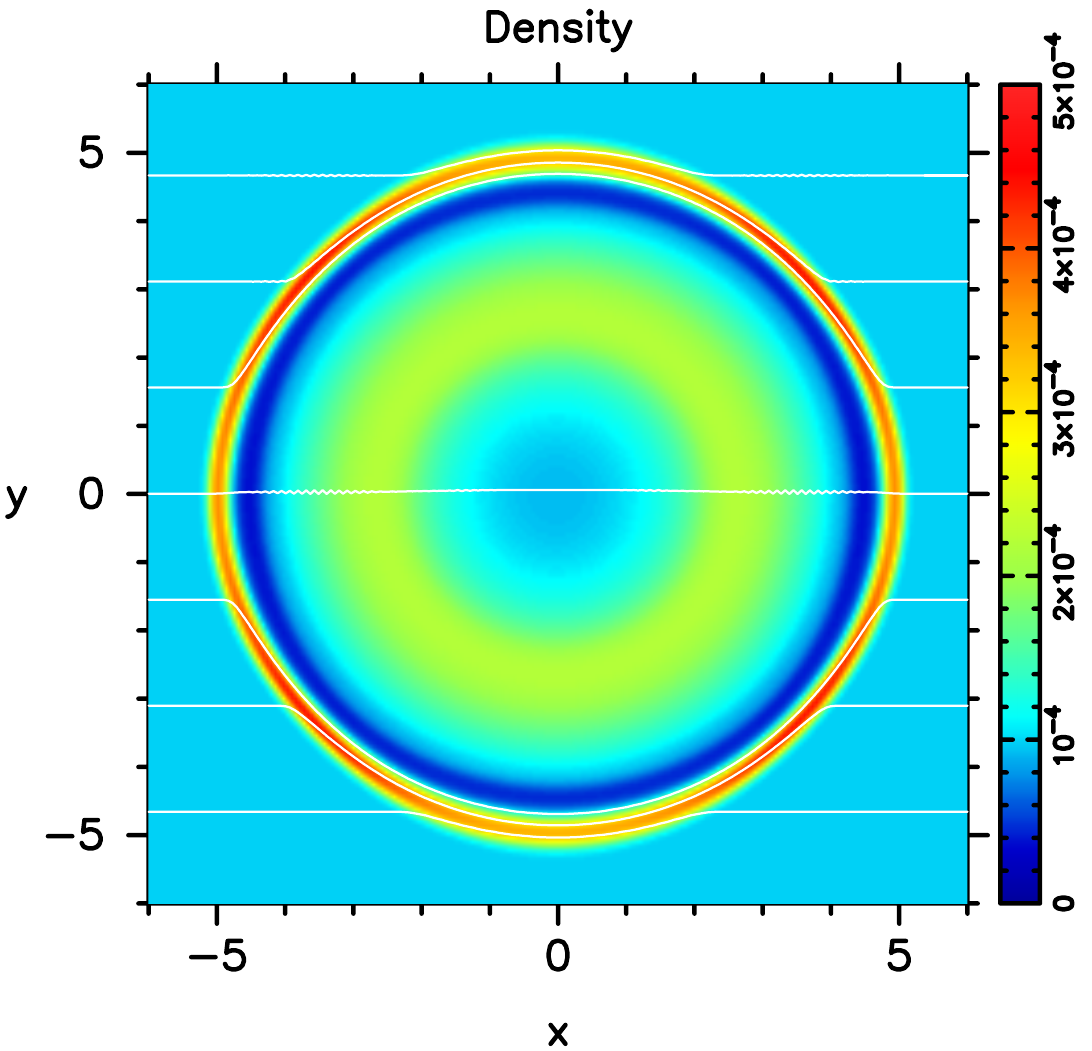}
\includegraphics[width=0.5\textwidth]{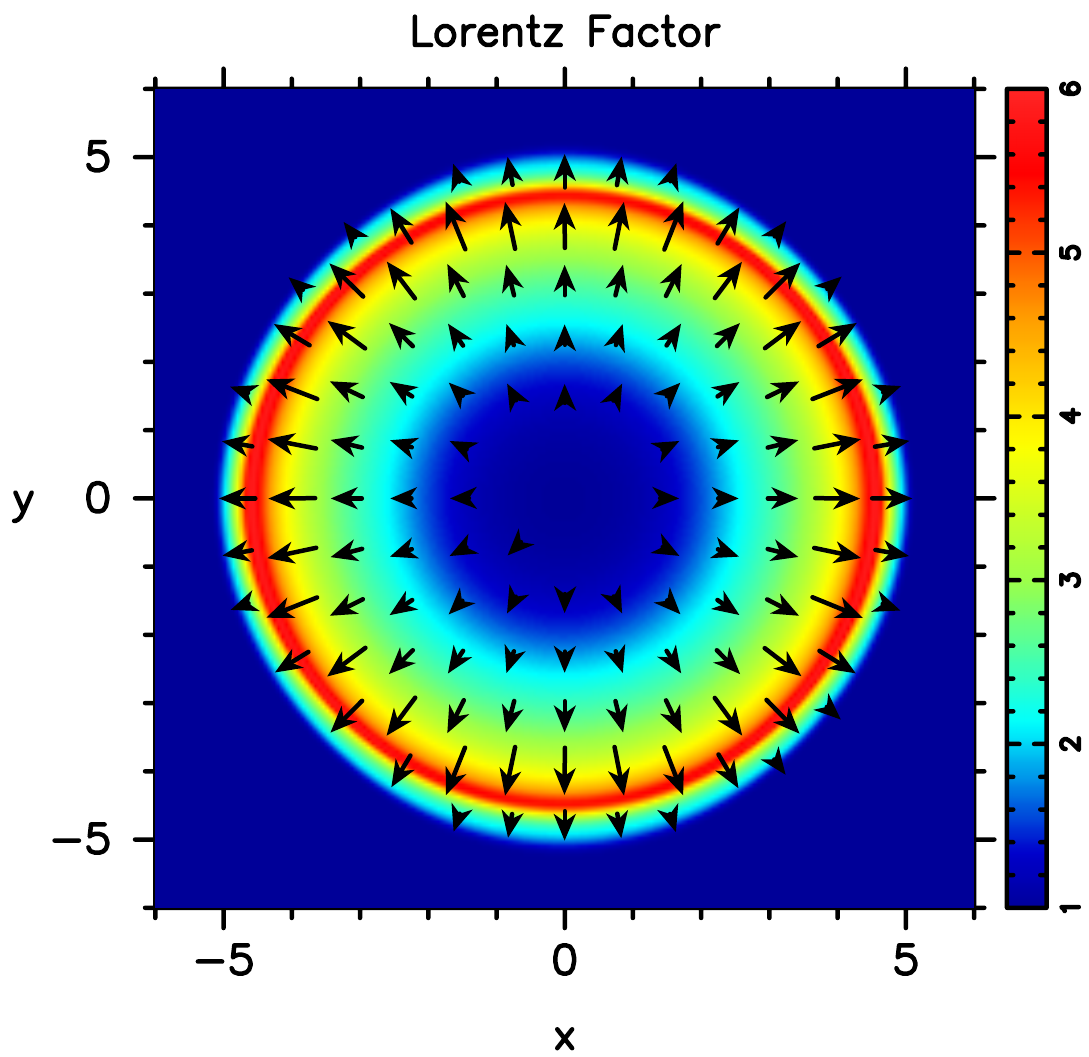}
\includegraphics[width=0.5\textwidth]{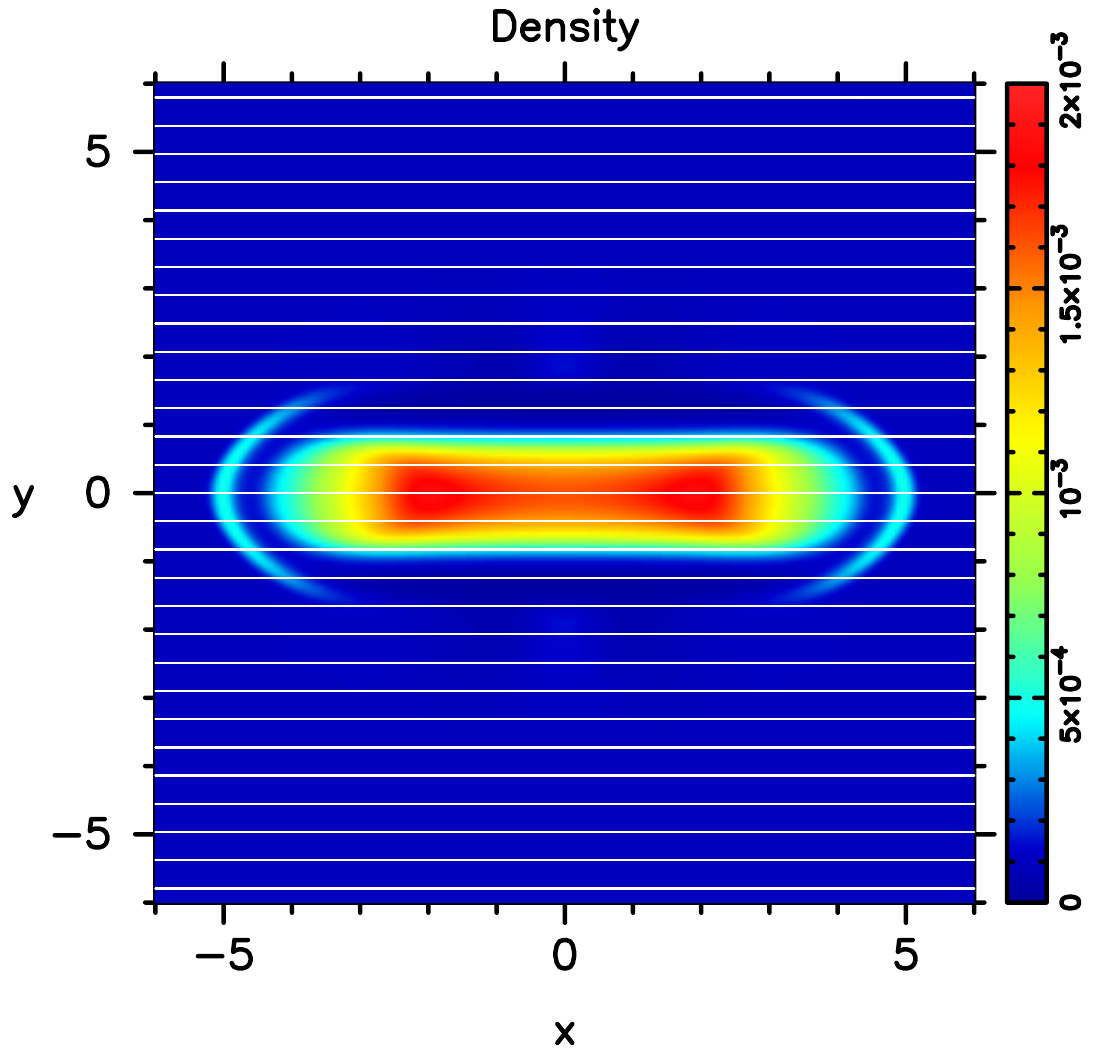}
\includegraphics[width=0.5\textwidth]{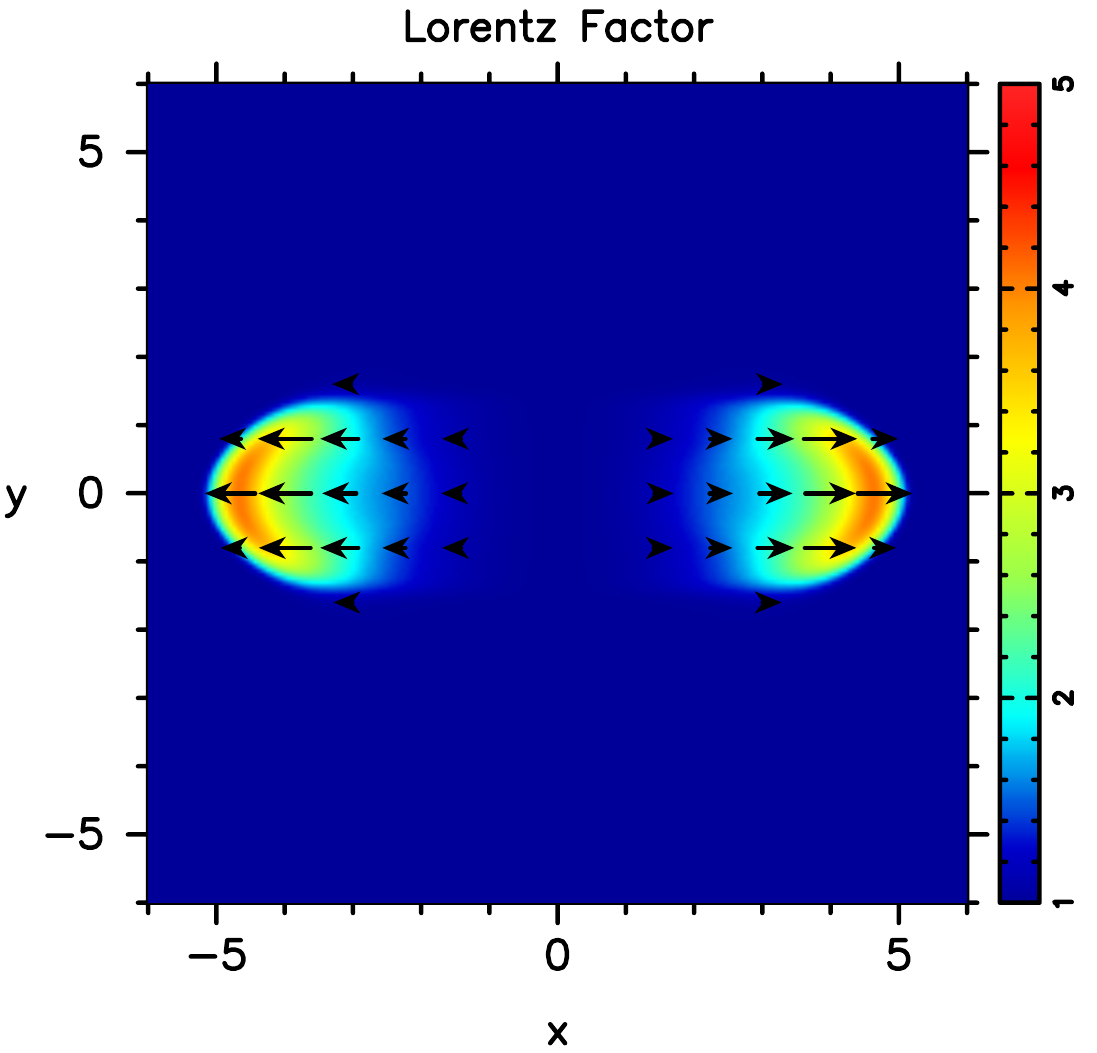}
\caption{\label{fig:CE}Cylindrical explosion test. The top panels show the solution for $\sigma_0=2.5\times10^{-5}$, whereas the bottom panels the one for $\sigma_0=2.5\times10^{5}$. In both cases, the integration time $t=4$.}
\end{figure}

\subsection{2D cylindrical explosion} 

This problem is set on a uniform Cartesian grid with 200 cells in each direction. 
The computational domain is $(-6,6)\times(-6,6)$. The initial solution describes a uniform ambient gas at rest with $p_a=3\times10^{-5}$, $\rho_a=10^{-4}$, $\vv{B}=(B_0,0,0)$, and $\vv{v}=(0,0,0)$. In the central region with the polar radius $r<0.8$ the gas pressure and density are raised to $p_0=1$ and $\rho_0=0.01$ respectively.  A $\tanh$-profile is used to introduce a gradual transition between the two states in $0.8<r<1.0$. This is exactly the same setup as in \cite{SSK99}, where $B_0=1$ was the highest value the code could handle (Later, with a somewhat modified code, the author of that paper failed to reproduce this result, as the code crashed.). We have had successful runs for $B_0=0.01,0.1,1,10,100$, and $1000$.  For $B_0=0.1$ the results look very close to that in \cite{SSK99} and other papers repeating this test (e.g. \cite{DelZanna07}).  Figure \ref{fig:CE} illustrates the results for the extreme cases of $B_0=0.01$ and $B_0=1000$; the corresponding magnetisation in the centre of the explosion, $\sigma_0=B_0^2/w(p_0,\rho_0)$, being $\sigma_0=2.5\times10^{-5}$ and  $\sigma_0=2.5\times10^{5}$ respectively.            

For $B_0=0.01$, the magnetic field has a very small effect on the flow up to $t=4$, and so it is almost the same as in the case of unmagnetised fluid. The magnetic field is swept aside by this flow. For $B_0=1000$, the magnetic field is so strong that the magnetic field lines are hardly deflected by the explosion and the fluid velocity is perfectly aligned with the x-axis.

\begin{figure}[h]
\includegraphics[width=0.5\textwidth]{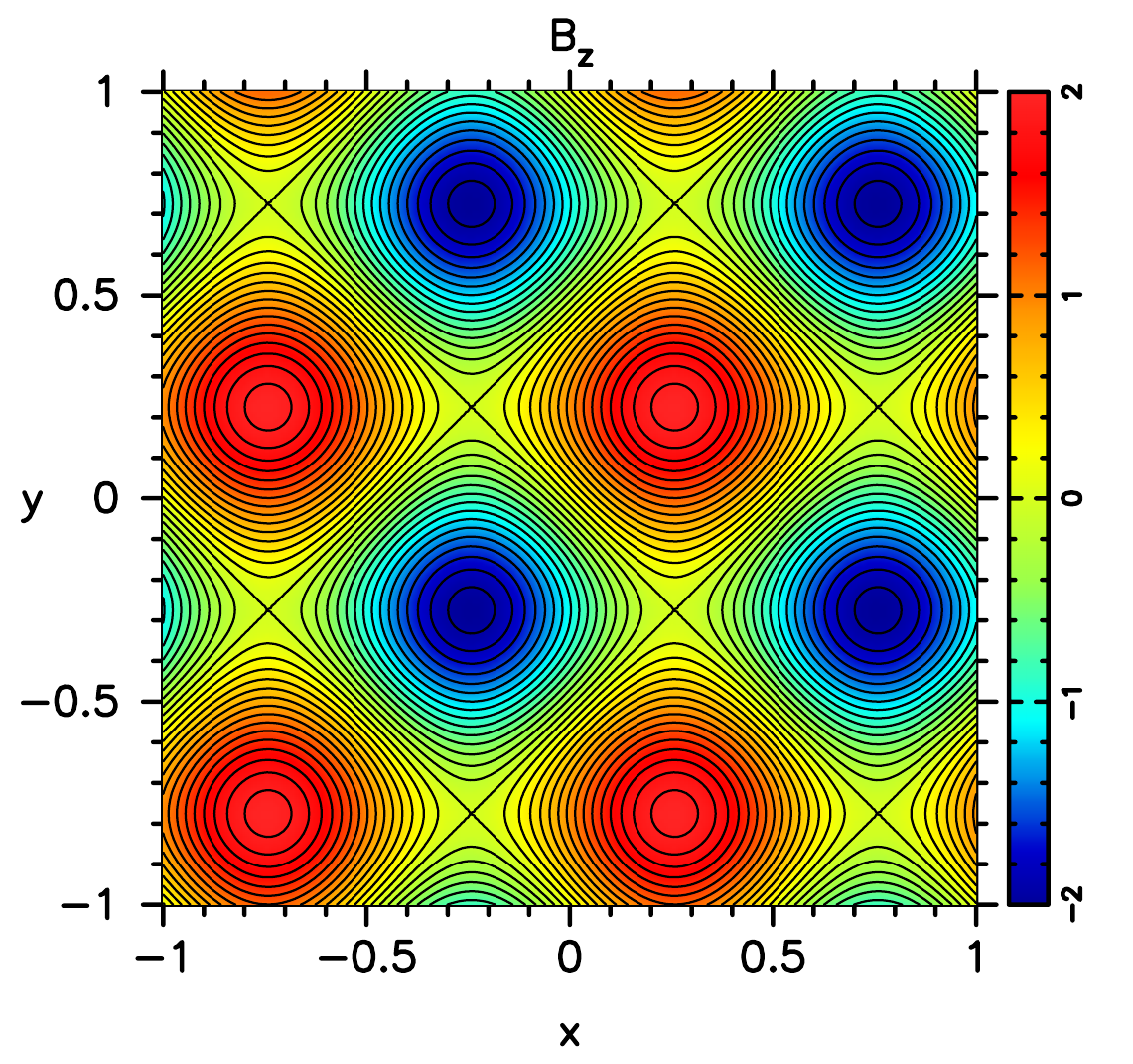}
\includegraphics[width=0.5\textwidth]{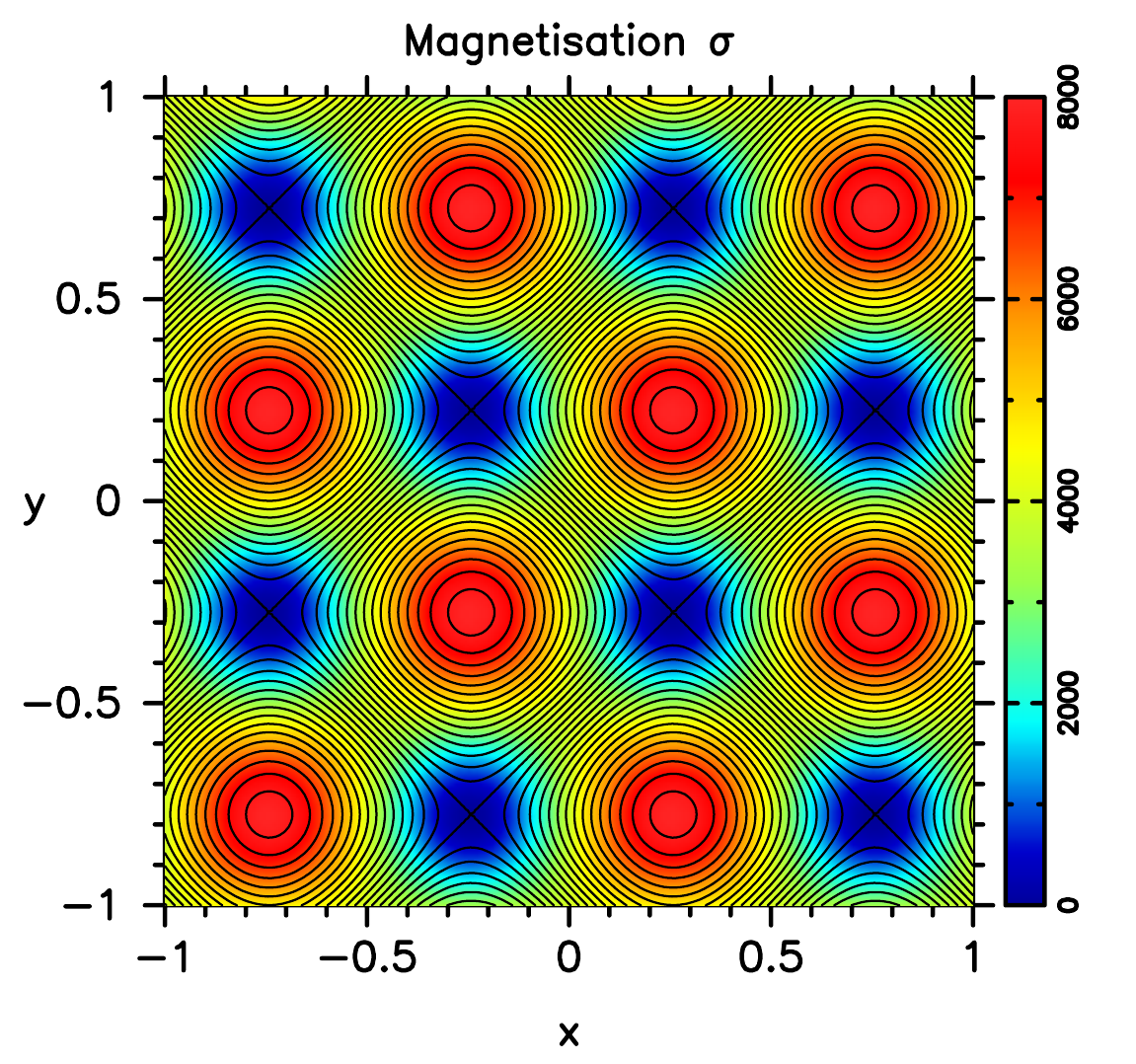}
\includegraphics[width=0.5\textwidth]{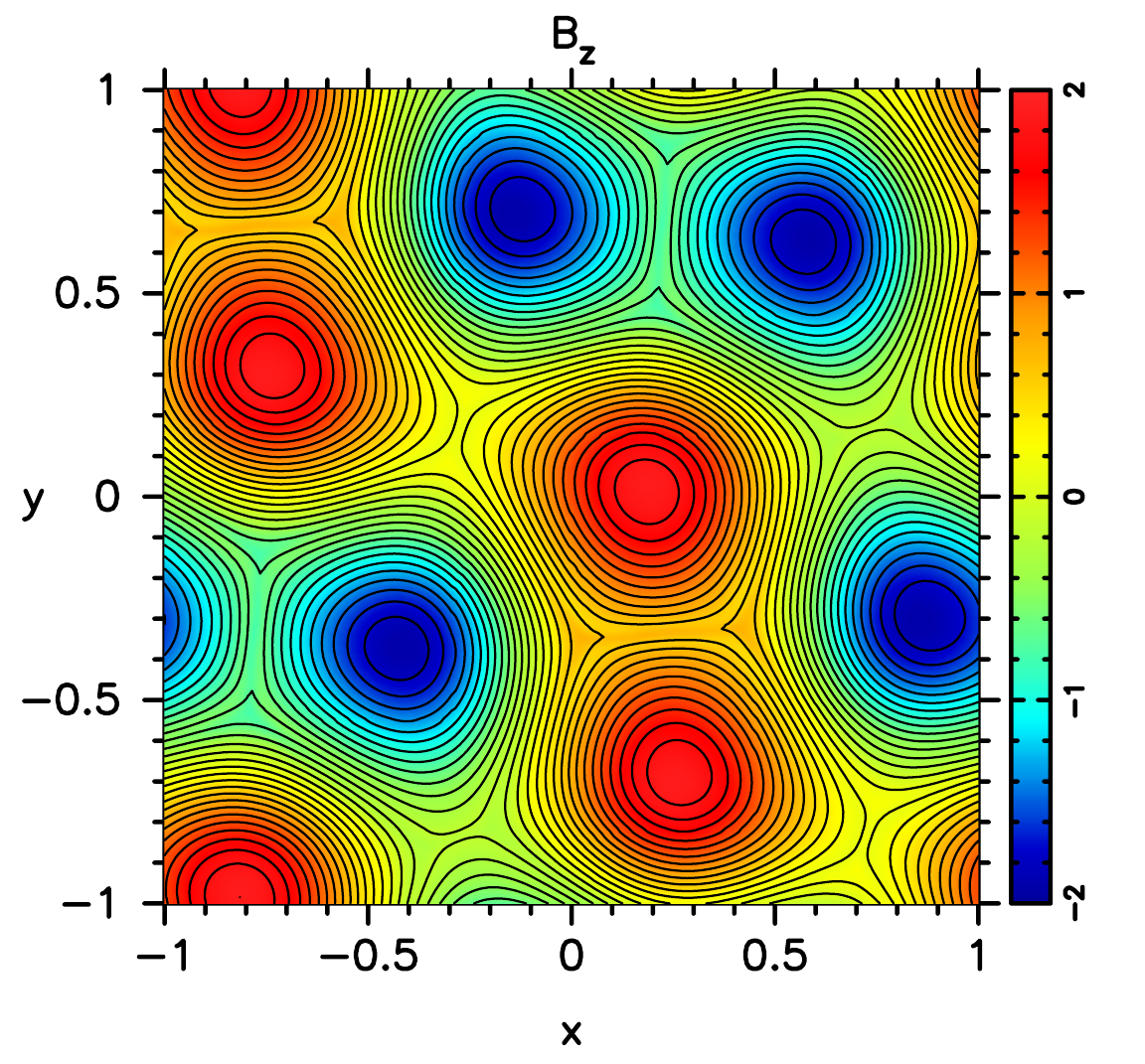}
\includegraphics[width=0.5\textwidth]{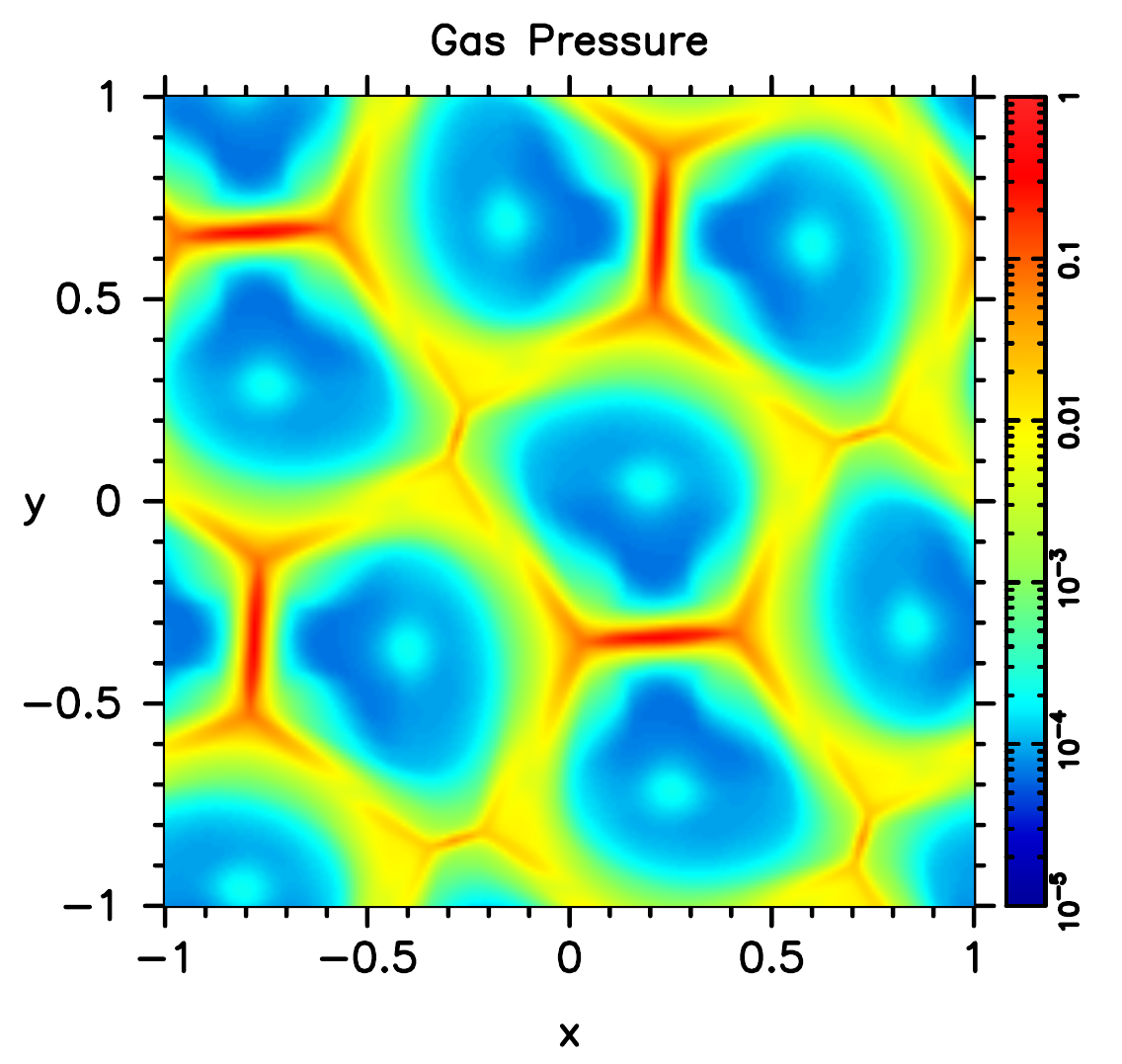}
\caption{\label{fig:MI} Collapse of the two-dimensional ABC grid. The top panels show the magnetic field lines and $B_z$ (left) and $\sigma$ (right) in the initial solution. The bottom panels show the solution at the onset of rapid merger of magnetic ropes with the same polarity ($t=7$): magnetic field lines and $B_z$ (left) and $p$ (right). }
\end{figure}

\subsection{2D ABC magnetic grid}

The initial configuration of this problem describes a static uniform plasma distribution of density $\rho_0$ and pressure $p_0$ and the periodic force-free magnetic field 

$$
\begin{cases}
B_x=-B_0 \sin(2\pi y)\,,\\
B_y=B_0 \sin(2\pi x)\,,\\
B_z=B_0 (\cos(2\pi x)+\cos(2\pi y))\,,
\end{cases}
$$ 
which describes an equidistant grid of twisted magnetic field ropes aligned with the z-axis. In the xy-plane,  these ropes appear as magnetic islands of opposite polarities (see the top-left pane of figure \ref{fig:MI}). Closest islands of the same polarities are separated via an x-point.  This equilibrium configuration is unstable, leading to the eventual merger of islands with the same polarity via a collapse of the x-points into current sheets, followed by magnetic reconnection. Here we aimed to reproduce the FFDE and PIC simulations of this process described in \cite{Lyutikov17}. 

The magnetic reconnection is accompanied by magnetic dissipation. In conservative schemes for RMHD, the total energy is conserved, and the energy of dissipated magnetic field is converted into heat.   This is not true for numerical FFDE, where the energy of dissipated magnetic field is lost. When no measures are taken to combat this energy loss, the total energy conservation in this test problem is strongly violated. The approach we used to overcome this problem involves 1) integrating the energy equation of FFDE sub-system together with the momentum equations, 2) computing the difference between the energy obtained this way and the energy computed from the $\vv{E}$ and $B$ fields of the sub-system, and 3) if it is positive, adding it to the energy of the perturbation sub-system at the end of every timestep. 

The parameters of the simulations described here are $B_0=1.0$, $p_0=\rho_0=10^{-4}$. The Cartesian computational domain is $(-1,1)\times(-1,1)$, with periodic boundary conditions and a $200\times200$ uniform grid.  Random noise was used to trigger the instability. 

\begin{figure}[h]
\includegraphics[width=0.5\textwidth]{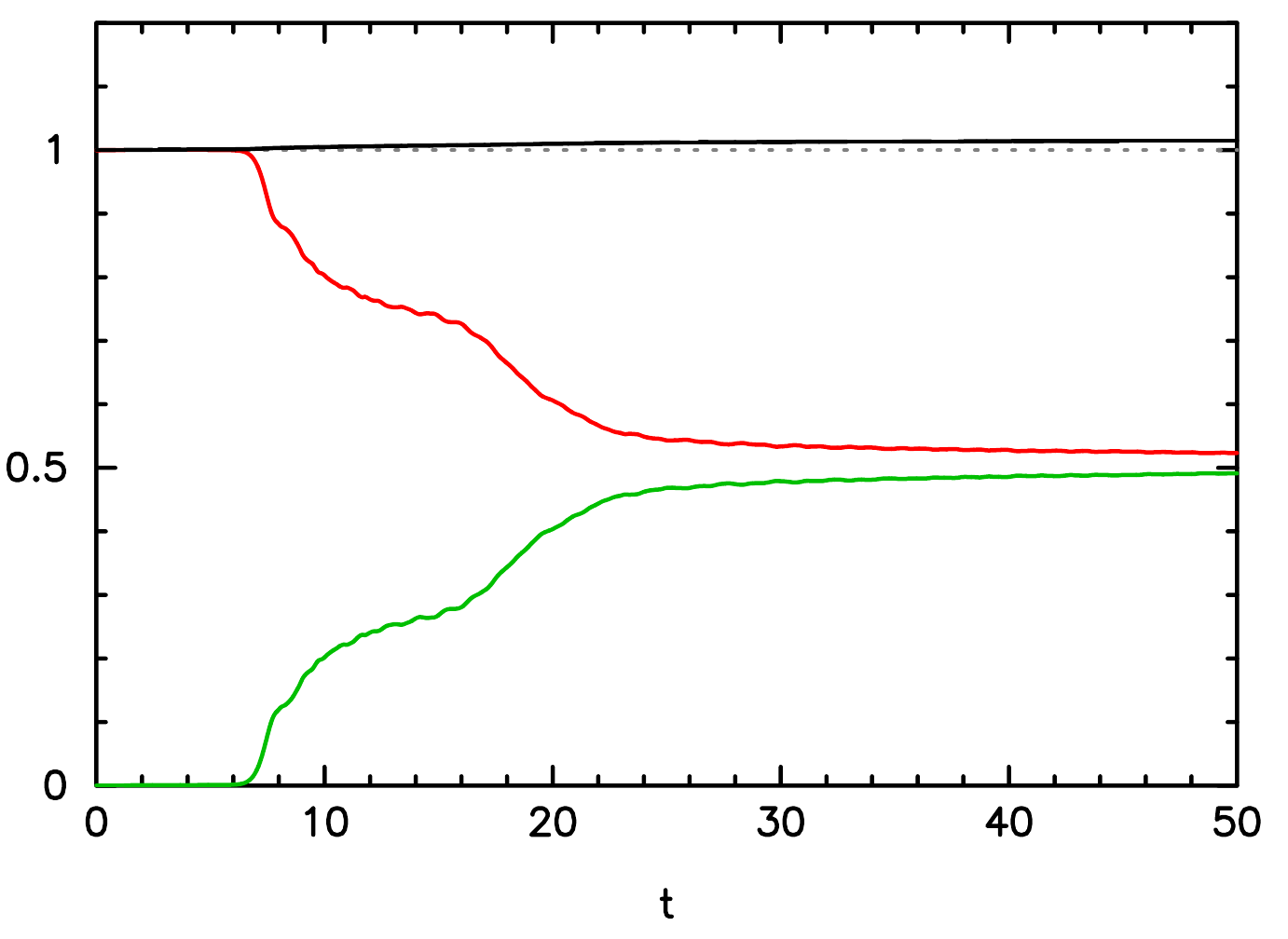}\hspace{2pc}%
\raisebox{3pc}{\begin{minipage}[b]{0.4\textwidth}\caption{\label{fig:ED} Energy balance in the ABS test. The black solid line shows the total energy in the box in the units of the initial total energy. The red line is the fraction of energy in the electromagnetic  field and the green line is the fraction of energy in the plasma.}
\end{minipage}}
\end{figure}

In the test simulations, the linear phase of the instability continues up to $t=6$. This phase corresponds to the initial plateau of the magnetic energy curve in figure \ref{fig:ED}. By $t=7$, the x-points separating the islands collapse into current sheets (see the bottom panels of figure \ref{fig:MI}). This opens the phase of rapid merger of neighbouring magnetic islands accompanied by strong magnetic dissipation.   Once the number of the islands in the box reduces to two (of opposing polarities) the dissipation rate slows down significantly. By $t=30$ almost 50 percent of the initial magnetic is converted into heat.   The timescales of the evolution are similar to that seen in the PIC simulations (see figure 9 in \cite{Lyutikov17}). Although we observe a  somewhat smaller fraction of the dissipated magnetic energy, this could be related to the four-times larger size of the computational box in the PIC simulations, allowing it to trace the mergers to larger length scales.  The total energy in the periodic box only slightly increases during the simulations.      

\section{Conclusions}

In this contribution, we have described a rather unusual version of the operator-splitting method tailored to overcome the stiffness of RMHD equations in the regime of ultra-high magnetisation.  The results of test simulations show that this approach allows to obtain accurate solutions for a wide range of problems involving smooth flows, shock waves, and magnetic reconnection.  It is suitable not only for highly magnetised flows but also for flows with low magnetisation, which is important for many astrophysical applications.   

\section*{References}
\bibliography{PhillipsKomissarov}
\end{document}